\documentclass[11pt,twoside]{article}
\usepackage{asp2010}
\resetcounters

\begin{document}

\title{The Odd Meanderings of the IMF Across Cosmic Time}
\author{Romeel Dav\'e$^1$}
\affil{$^1$Astronomy Department, University of Arizona, Tucson, AZ 85750}

 \begin{abstract}
It is difficult to reconcile the observed evolution of the star
formation rate versus stellar mass (SFR-$M_*$) relation with
expectations from current hierarchical galaxy formation models.
The observed SFR-$M_*$ relation shows a rapid rise in SFR($M_*$)
from $z=0\rightarrow 2$, and then a surprisingly lack of amplitude
evolution out to $z\sim 6+$.  Hierarchical models of galaxy formation
match this trend qualitatively but not quantitatively, with a maximum
discrepancy of $\sim\times 3$ in SFR at $z\sim 2$.  One explanation,
albeit radical, is that the IMF becomes modestly weighted towards
massive stars out to $z\sim 2$, and then evolves back towards its
present-day form by $z\sim 4$ or so.  We observe that this redshift
trend mimics that of the cosmic fraction of obscured star formation,
perhaps hinting at a physical connection.  Such IMF evolution would
concurrently go towards explaining persistent discrepancies between
integrated measures of star formation and present-day stellar mass
or cosmic colors.
\end{abstract}

\section{Introduction}

There is currently much debate on whether the stellar initial mass
function (IMF) is everywhere and everywhen invariant, as highlighted
by this excellent conference.  Direct evidence for IMF variations
are scant and controversial~\citep[and contributions to these
proceedings by Meurer, Lee, and Hoversten]{bas10}.  One possibility
is that the cluster-intrinsic IMF is invariant, but sampling and/or
truncation effects convolved with a varying cluster mass distribution
yields a varying IMF when averaged over entire galaxies (see e.g.
contributions on the IGIMF theory).  Theoretically it is difficult
to envision that, despite enormous variations in the physical
conditions of star formation, the IMF somehow conspires to remain
completely invariant~\citep{kro02}.  But the sense in which the IMF
should vary is not well understood, and hence more empirical
constraints are needed.

Recently, a cosmological approach to constraining the IMF has gained
traction.  Here, the evolutionary properties of galaxies are studied
to test whether they are consistent with an invariant IMF.  Early
efforts involved comparing the observed integral of cosmic star
formation (which traces massive star evolution) with the present-day
stellar mass density (which traces $\sim 1M_\odot$ stars).  Such
studies typically favored a cosmically-averaged IMF that was weighted
towards more massive stars than the present-day Milky Way disk
IMF~\citep{mad98,bal03,wil08}.  Other approaches include studying
the color evolution of passive galaxies~\citep{vand08}, and applying
additional constraints from extragalactic background light
measures~\citep{far07}.  In general, such studies again favor an
IMF more weighted towards massive stars to satisfy all constraints.
That said, a well-known theorem~\citep{bas10} states that any problem
in galaxy formation can be resolved through a suitable choice of
IMF, and hence such arguments are often dismissed as the last resort
of a scoundrel.  This fails to deter us theorists, who have been
called far worse.

In \citet{dav08}, we argued that the evolution of the star formation
rate-stellar mass (SFR$-M_*$) relation may be suggestive of a
time-varying IMF for typical star-forming galaxies.  At that time,
reliable observations of SFR$-M_*$ only extended out to $z\sim 2$.
Improved data has now constrained SFR$-M_*$ to $z\sim
7$~\citep{sta09,lab10,gon10}, providing an opporunity to extend
such arguments to very high redshifts.  Concurrently, {\it Herchel}
has provided improved bolometric constraints on star formation in
galaxies out to $z\sim 2$, mitigating some of the systematic
uncertainties that may have affected earlier works.  Theoretical
work has also progressed, as toy models have been developed that
can explain SFR$-M_*$ evolution~\citep{bou10}, at the cost of
introducing arbitrary modifications to galaxy star formation
histories.  With such fervent recent activity, the time is right to revisit
our previous arguments.  This is the goal of these proceedings.
The end result, we argue, is that the evidence still favors some
discrepancy in the SFR vs. $M_*$ evolution in galaxies that could
be resolved by a time-varying IMF, though to do so requires a
modestly varying IMF that evolves non-monotonically with redshift.

\section{The SFR$-M_*$ Relation}

The SFR$-M_*$ relation has recently received much attention as an
important barometer of galaxy evolution theory.  Models, particularly
cosmological hydrodynamic simulations, have long predicted that
there should be a tight and slightly sub-linear relation between these
quantities~\citep[e.g.][]{dav00,fin06,dav06} that evolves slowly
upwards with redshift~\citep{dav08}.  Observations over the last
five years have confirmed these trends out to $z\sim
1$~\citep{noe07,elb07}, $z\sim 2$~\citep{dad07}, and beyond~\citep{gon10}.
In models, this trend arises because cold and smooth accretion is
the dominant accretion path for fueling star formation at all
epochs~\citep{mur02,ker05,dek09}; the tightness of SFR$-M_*$
empirically constrains the contribution from
starbursts~\citep[e.g.][]{noe07}.  The broad agreement between
theory and observations is a key success of current galaxy formation
models.

\begin{figure}
\vskip-0.4in
\plotone{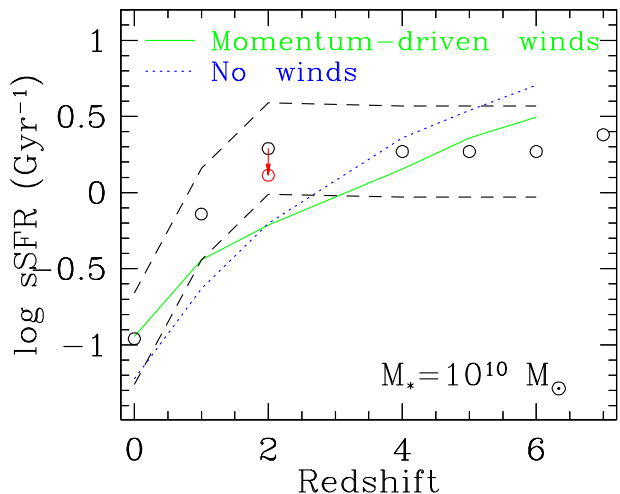}
\vskip-3.6in
\caption{The amplitude evolution of the SFR$-M_*$ relation at 
$M_*=10^{10} M_\odot$ from $z=0-7$.  Observations (circles) are shown 
from \citet{elb07} at $z=0,1$, and from \citet{gon10} at $z\geq 2$.
Dashed lines indicate $\pm0.3$~dex spread around the mean relation.
Red dot roughly indicates the downward correction
of UV-derived SFRs based on {\it Herschel} data.
Simulation results are shown for our
favored run employing momentum-driven winds (green solid)
and a run with no galactic outflows (blue dotted).
\label{fig:sfrmstar}
}
\end{figure}

In detail, however, it has been noted that SFR$-M_*$ predicted in
models (either hydrodynamic or semi-analytic) fails to match the
observed amplitude evolution~\citep{elb07,dad07,dav08}.  This is
shown in Figure~\ref{fig:sfrmstar}, which shows the evolution of
the SFR$-M_*$ relation at a fiducial stellar mass of $M_*=10^{10}
M_\odot$ from two cosmological hydrodynamic simulations~\citep[details
of which are in][]{opp10}.  The green curve shows results from a
simulation including outflows with scalings as expected for
momentum-driven winds; this outflow model has proved successful at
matching a wide range of galaxy and IGM data from $z\sim 0-6$.  The
blue curve shows a simulation with no galactic outflows included,
which strongly overpredicts the global stellar mass formed~\citep{opp10},
but nevertheless produces a SFR$-M_*$ relation that is not markedly
different from the strong outflow case.  This illustrates that
despite dramatic variations in predictions owing to different
feedback models, the SFR$-M_*$ relation is relatively insensitive
to feedback~\citep{dav08,dut10}.  The basic reason is that feedback
suppresses star formation which likewise suppresses stellar mass
growth, moving galaxies downwards along the (nearly linear) SFR$-M_*$
relation.
%The curves are obtained
%by fitting a power law to the simulated galaxies from $\approx
%10^9-10^{11}M_\odot$, although the simulated SFR$-M_*$ relation
%does not strictly follow a single power law over the entire range.

Observations from \citet[at $z=0,1$]{elb07} and~\citet[$z\geq 2$]{gon10}
are shown as circles,
and a typical $1\sigma$ observed scatter around the relation of $\pm0.3$ dex
is shown by the dashed lines.  Note that this is {\it not} the
statistical or systematic uncertainties in $M_*$ or SFR, which are
typically smaller than 0.3~dex at low-$z$ but could be significantly
larger at high-$z$.  As can be seen, the observed SFR$-M_*$ relation
amplitude matches well at $z\sim 0$, and then the observed SFRs are
increasingly larger going out to $z\sim 2$.  At this point, there
is an abrupt change in the trend, and the observed amplitude is
essentially invariant out to $z\sim 7$.  The simulations broadly
predict a similar trend, but do not reproduce the abrupt change at
$z\sim 2$.

The discrepancies are modest, especially considering the uncertainties
involved, maximizing at $\sim\times 3$ at $z\sim 2$.  But what is
particularly puzzling is that at $z\sim 2$, the observed star
formation rates can exceed the accretion rate into galaxy halos,
which is difficult to accomodate in current models that favor
continual gas supply (``cold streams") to fuel star formation.  The
implication is that galaxies must store up large reservoirs of gas,
and then something triggers its rapid conversion into stars.  If
star formation proceeds exponentially, then galaxies quickly move
upwards along the SFR$-M_*$ relation~\citet{mar10}, preserving the
(nearly) linear slope at a fixed amplitude.  Current hierarchical
galaxy formation models do not yield this behavior naturally.  \citet{bou10}
noted that an ad hoc way to accomplish this is to explicitly prevent
all star formation in halos with $M<10^{11}M_\odot$, which accumulates
a large reservoir that is then rapidly consumed once the threshold
halo mass is crossed.  But this is difficult to arrange physically,
and the lack of any discernible transition in the galaxy population
around these halo masses (as compared to, say, around $10^{12}M_\odot$)
argues against this scenario.  Hence matching the $z\sim 2+$ SFR$-M_*$
relation appears to require a fundamental modification to a galaxy
formation paradigm that is otherwise quite successful at reproducing
a wide range of observations~\citep[e.g.][]{ben10}.

An alternative possibility is that there are significant systematics
in observational measures of SFR and/or $M_*$.  As discussed in
\citet{dav08}, it is easier to accommodate overestimated SFR's than
underestimated $M_*$, since stellar mass is cumulative and there
are lower-$z$ constraints that must be satisfied.  The \citet{dad07}
$z\sim 2$ data rely on extinction-corrected UV SFRs, which recent
{\it Herschel} data suggest overestimate the true (bolometric) SFRs
measured from far-IR data by 50\% or perhaps up to $\sim\times 2$.
Such a reduction is indicated by the red point on
Figure~\ref{fig:sfrmstar}.  This would alleviate the discrepancy
somewhat but not completely, and in a sense makes it harder to
reconcile since a key systematic has been mitigated.  We note that
24$\mu$-based SFRs have been shown to be very poor at $z\sim 2$
since PAH emission that dominates the rest-frame 8$\mu$ emission
is poorly understood~\citep[e.g.][]{kel10}, but such SFRs are not the basis for the
SFR$-M_*$ observations shown.  Nevertheless, better data is required
to pin all the systematics down, and a proper accounting may yet
remove the discrepancy altogether.

Finally, Figure~\ref{fig:sfrmstar} shows that feedback does have a
noticeable impact on the shape of SFR$-M_*$ evolution.  In particular,
the momentum-driven wind scalings suppress star formation in smaller
halos more, resulting in less early star formation in better agreement
with data.  One might envision that an even stronger feedback model
could completely reconcile the theory and observations.  However,
the model of \citet{bou10} indicates that the feedback required is
quite extreme, essentially suppressing all star formation even in
fairly sizeable halos.  Our experience is that models with significantly
stronger feedback have substantial difficulties producing enough
early star formation to match data on $z\sim 6+$ galaxies, the
mass-metallicity relation, and the metallicity of the IGM at early
epochs.  Perhaps some compromise will eventually be found between
theory and observations, and certainly the downward revision of
observed $z\sim 2$ SFRs goes towards this.  This probably remains
the most likely solution to the SFR$-M_*$ discrepancy, but it is
not the most interesting one.

\section{A Time-Varying IMF?}

A more interesting and far-reaching systematic that could lower
inferred SFRs is if the IMF is weighted towards more massive stars
at high redshift, either by being top-heavy or bottom-light.  Here
I use top-heavy to indicate a changing upper-end slope, and
bottom-light to refer to the suppression of low-mass stars.  If
this is to be the solution, then the IMF must evolve with redshift
towards more top-heavy/bottom-light at $z\sim 2$, and then {\it
reverse} its trend back towards a normal IMF by $z\sim 4+$.  At
$z\ga 5$ the systematic effects in $M_*$ and SFR are currently
sufficiently large that it would be premature to infer anything
from the small discrepancies at those redshifts.

\begin{figure}
\vskip-0.4in
\plotone{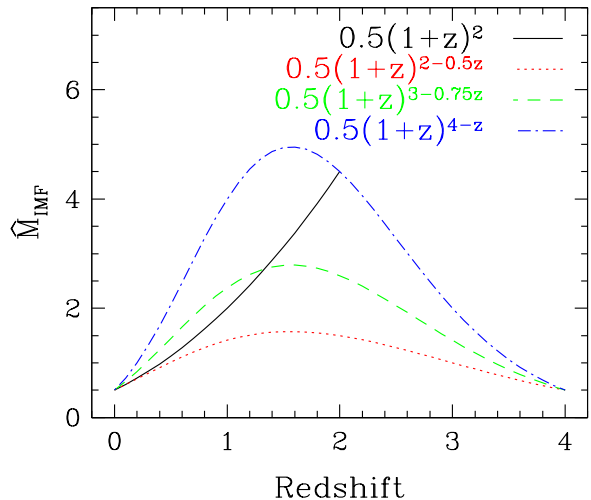}
\vskip-3.6in
\caption{The IMF characteristic mass $\hat{M}_{\rm IMF}$
in various IMF evolutionary forms.  The solid black line shows
the form of \citet{dav08}, which reconciles SFR-$M_*$ from $z=0\rightarrow 2$.  
Recent results have suggested the IMF returns
to its present-day form by $z\sim 4$, motivating a different
evolutionary form.  The green dashed line, with
$\hat{M}_{\rm IMF}=0.5(1+z)^{3-0.75z}\;M_\odot$, follows the
old form out to $z\sim 1.5$, and then returns
towards the present-day IMF by $z=4$, broadly accommodating
current constraints.
\label{fig:mc}
}
\end{figure}

One possible form of IMF variation, highlighted in \citet{dav08}
and \citet{vand08}, is that the IMF characteristic mass evolves
with redshift.  For a Kroupa IMF, i.e.  $\frac{dN}{d\log{M}} \propto
M^{-0.3}$ for $M<\hat{M}_{\rm IMF}$ and $\frac{dN}{d\log{M}} \propto
M^{-1.3}$ for $M>\hat{M}_{\rm IMF}$, \citet{dav08} showed that one
evolutionary form that reconciles SFR-$M_*$ evolution out to $z\sim
2$ is $\hat{M}_{\rm IMF}=0.5 (1+z)^{2} M_\odot$.  However, beyond
$z\sim 2$ the IMF must evolve back towards the present-day form,
and moreover Herschel data has shifted the $z\sim 2$ point downwards.
A general form that would accommodate these trends is
\begin{equation}
\hat{M}_{\rm IMF}=0.5(1+z)^{3-0.75z}\;M_\odot 
\end{equation}
from $z=0\rightarrow 4$, remaining at $0.5 M_\odot$ at $z\geq 4$.
As shown in Figure~\ref{fig:mc}, this roughly mimics the $\hat{M}_{\rm
IMF}$ evolution of \citet{dav08} out to $z\sim 1.5$, and then turns
over, reducing the characteristic mass by almost a factor of two
at $z\sim 2$, and eventually returns to the present-day value at
$z=4$.  This is purely an eyeball estimate, and furthermore this
choice of redshift dependence is merely a convenient parameterization;
there is no physics in it.  In reality, the IMF should be tied to
some physical property of galaxies that evolves with redshift, such as star formation
rate or surface density.

The SFR$-M_*$ relation does not constrain the detailed
form of IMF variation.  Other plausible ways to reconcile the
SFR$-M_*$ would be to make the upper-end slope of the IMF more
shallow with time~\citep[e.g.][]{wil08}.  In detailed studies of
lensed Lyman Break ($z\sim 2-3$) galaxies, the upper end of the IMF
does not appear to be substantially different~\citep{qui09} though
current constraints may not rule out mild changes that may resolve
the SFR$-M_*$ discrepancy.  Also, constraints on the IMF in
$z\sim 2$ sub-millimeter galaxies indicate that it is consistent
with a \citet{cha03} IMF, perhaps slightly favoring a more top-heavy
one, but clearly inconsistent with dramatic departures from a
Salpeter slope as postulated in e.g. the semi-analytic models of
\citet{bau05}.  On the other hand, the ratio of H$\alpha$ to UV
emission in local star-forming galaxies correlates strongly with
its star formation rate and/or surface density~\citep{meu09,lee09},
for which one interpretation is that the IMF has a shallower upper-end
slope where there is more concentrated star formation.  Since $z\sim
2-3$ galaxies ubiquitously show more concentrated star formation,
this could empirically imply a top-heavy IMF in such systems.  It
remains to be seen if simply applying constraints from local galaxies
would yield the required evolution owing purely to the size and
star formation rate evolution of typical galaxies, but this should
be investigated.

An IMF that returns to normal at very high redshifts is also preferred
by available constraints at those epochs.  For instance, current
models assuming a normal IMF are able to form enough stellar mass
in very early ($z\sim 6+$) objects as observed~\citep[e.g.][]{fin10},
but a significantly bottom-light or top-heavy IMF would create
difficulties.  Another early-universe constraint comes from old
globular clusters formed at high-$z$, which are consistent with a
present-day IMF.  Hence there is a growing concensus that the IMF
in the early universe is actually quite similar to today's.  Of
course, Population III (i.e.  very metal poor) stars are expected
to have a quite top-heavy IMF, but it is unlikely that these
contribute significantly to early galaxies observed to
date~\citep[e.g.][]{dav06,fin10}.

The central question is, what could cause the IMF to depart away
from the present-day IMF in such a fashion?  The non-monotonic
evolution is particularly challenging to understand.  This is a
more a question for star formation theorists, but there is one
possible intriguing connection:  The required IMF evolution roughly
mimics the cosmic fraction of obscured star formation as determined
by \citet[see their Figure~11]{bou09}, which peaks at $z\sim 2-3$
and falls off to higher and lower redshifts.  Hence if there was
something about an obscured mode of star formation that systematically
caused the IMF to be weighted towards massive stars, this may yield
the desired trend.  A possible local test would then be to look in
nearby ULIRGs, where some controversial claims of a top-heavy IMF
have been made~\citep{rie80,rie93}.  Obviously this connection is
highly tenuous, but may provide an avenue for further exploration.

\section{Cosmically-Averaged Star Formation}

The IMF variation required to reconcile theory and observations of
SFR$-M_*$ also broadly reconciles the observed cosmic star formation
and stellar mass growth histories~\citep{dav08,wil08}, and extragalactic
background light constraints~\citep{far07}.  Clearly these quantities
are all interrelated, so it isn't terribly surprising that the same
IMF fixes all of them, but it provides a useful consistency check.
Another approach (see Wilkins, these proceedings) is to use the
cosmic SED to constrain the IMF; once again, this shows a mismatch
between the integrated cosmic star formation history and the
present-day mean cosmic color in the analogous sense that there is
too little near-IR light relative to the UV/optical light together
with expectations from the observed cosmic star formation history.
Present-day integrated measures do not necessarily require an {\it
evolving} IMF, but rather only a cosmically-averaged IMF that differs
from the local one.  However, it would be puzzling if our
locally-measured IMF was different than the (invariant) IMF over
the rest of cosmic time.

Conversely, differential measures of stellar mass growth and star
formation rate could imply an evolving IMF, and some claims to that
end have been made~\citep{wil08,dav08}.  However, such claims are
controversial.  A particularly strong argument was forwarded by
\citet[also see these proceedings]{red09}, who noted that by making
careful corrections for dust and incompleteness, it is possible to
reconcile the global star formation rate and stellar mass density
evolution at $z\sim 2$.  This would not counter arguments based
on individual galaxies such as the SFR$-M_*$ discrepancy, but would
make observations of cosmic star formation and mass growth
internally self-consistent at $z\sim 2$.

In actuality, however, this argument ends up merely pushing the
problem to lower redshifts.  To see this, \citet{red09} notes that
their incompleteness corrections result in 57\% of the present-day
stellar mass already having been formed by $z\sim 2$.  Conversely,
current observations of the cosmic star formation history imply
that only $\sim 30\%$ of total cosmic star formation occurs prior
to $z=2$~\citep[e.g.][]{hop06,far07} --- this is independent of
assumed IMF, so long as it is time-invariant.  Hence the growth
from $z=2\rightarrow 0$ of stellar mass does not match expectations
from the observed star formation history, in the sense (once again)
that star formation rates are measured to be too high.  Hence one
can ``fix" the problem at $z\sim 2$ and have difficulty with evolution
to $z=0$, or one can ``fix" the problem at $z=0$ and then have
difficulty reconciling $z\sim 2$ observations.  This argues for
systematics in SFR or $M_*$ measures that change with redshift, be
they related to the IMF or not.

\section{Conclusion}

There is still no definitive evidence for IMF variations across
cosmic time.  That said, the issues motivating serious consideration
of such variations have yet to be fully resolved.  New data on the
SFR$-M_*$ relation out to $z\sim 7$ suggests that, if IMF variations
are to explain the discrepancy with theoretical expectations,
deviations from the present-day IMF must maximize at $z\sim 2$ and
then lessen to higher redshifts, returning to a standard IMF by
$z\sim 4-5$.  A possible form that would roughly yield this is a
Kroupa IMF with a characteristic (turnover) mass evolving as
$\hat{M}_{\rm IMF}=0.5 (1+z)^{3-0.75z}\; M_\odot$.  While the
required non-monotonic IMF evolution is not straightforward to
understand physically, a quantity showing analogous behavior
is the cosmic fraction of obscured star formation, hinting at a
possible physical connection.

The form of the IMF variation remains difficult to pin down.  The
best (albeit controversial) evidence for local variation is in the
form of a changing upper-end slope with star formation rate.  The
high star formation rates of high-$z$ galaxies would then naturally
result in a more top-heavy IMF.  But direct observations of $z\sim
2-3$ galaxies disfavor significant changes in the upper-end IMF
slope.  Meanwhile, the evolutionary form suggested by
\citet{dav08}~\citep[or, similarly, by][]{vand08} towards more
bottom-light with redshift is more difficult to test directly.  It
is also possible that the intrinsic IMF is invariant, but the larger
typical sizes of molecular clouds in high-$z$ galaxies results in
a galactic-averaged IMF that yields proportionally more massive
stars.  An IMF weighted towards massive stars at intermediate
redshifts would also go towards resolving the persistent discrepancies
between integrated measures of cosmic star formation history and
present-day measures of stellar mass or cosmic light.

In the near future, surveys such as the {\it Spitzer} Extragalactic
Deep Survey (PI G. Fazio) and the {\it Hubble} CANDELS survey (PIs
S. Faber and H. Ferguson) will provide unprecedented multiwavelength
samples of galaxies out to $z\sim 6$ and beyond.  Although this
will not necessarily pin down systematic effects in SFR or $M_*$
measures, it will enable tests of different galaxy formation scenarios
with greater statistical precision.  For instance, the ``reservoir"
scenario where gas consumption proceeds exponentially in galaxies
predicts a significantly different evolution to high-$z$ of the
stellar mass function and halo occupancy than hierarchical galaxy
formation models.  Hence these surveys will be critical for testing
current galaxy formation models, thereby indirectly testing ideas such as
IMF evolution.  We look forward to engaging in such comparisons.

\acknowledgements The author would like to thank Ben Oppenheimer,
Kristian Finlator, Du\v{s}an Kere\v{s}, Neal Katz, Avishai Dekel,
and Nicolas Bouch\'e for helpful discussions. Also thanks to
\citet{up2010loc} for a stimulating meeting!

\bibliography{dave_r}

\end{document}